\documentclass[preprint,12pt]{elsarticle}

\usepackage{amssymb}

\usepackage{amssymb}

\usepackage{url}
\usepackage{caption}
\usepackage{subcaption}
\usepackage{amsmath}
\usepackage{float}
\usepackage{booktabs}
\usepackage{multirow}
\usepackage[dvipsnames]{xcolor}
\usepackage{siunitx}

\newcommand{\changedOne}[1]{#1}
\newcommand{\changedTwo}[1]{#1}
\newcommand{\changedBoth}[1]{#1}

\journal{Acta Astronautica}

\begin{document}

\begin{frontmatter}


\title{On Asteroid Retrieval Missions Enabled by Invariant Manifold Dynamics}

\author[add1]{Jack Tyler\corref{corrauthor}}
\ead{jack.tyler@soton.ac.uk}
\cortext[corrauthor]{Corresponding author}

\author[add1]{Dr. Alexander Wittig} 
\ead{a.wittig@soton.ac.uk}

\address[add1]{Astronautics Research Group, University of Southampton, Southampton, United Kingdom, SO17 1BJ}

\begin{abstract}
In recent years, the retrieval of entire asteroids has received significant attention, with many approaches leveraging the invariant manifolds of the Circular-Restricted Three-body Problem to capture an asteroid into a periodic orbit about the $L_1$ or $L_2$ points of the Sun-Earth system. Previous works defined an `Easily Retrievable Object' (ERO) as any Near-Earth Object (NEO) which is retrievable using these invariant manifolds with an impulsive $\Delta v$ of less than $500$ m/s. We extend the previous literature by analysing the Pareto fronts for the EROs discovered for the first time, using high-performance computing to lift optimisation constraints used in previous literature, and modifying the method used to filter unsuitable NEOs from the NEO catalogue. In doing so, we can demonstrate that EROs have approximately the same transfer cost for almost any possible transfer time, including single-impulse transfers, which could offer significant flexibility to mission designers. We also identify $44$ EROs, of which $27$ are new, and improve on previously-known transfer solutions by up to $443$ m/s, including $17$ new capture trajectories with $\Delta v$ costs of less than $100$ m/s.
\end{abstract}




\end{frontmatter}



\section{Introduction}
\label{sec:intro}

In recent years, there has been significant interest in the investigation and exploration of Near-Earth Objects (NEOs), asteroids and comets which pass within 1.3 Astronomical Units (AU) of the Sun. Much of this is motivated by their composition, which mirrors the structure of the early Solar System and thus offers unique, invaluable insight into the history of our stellar neighbourhood \cite{Zimmer2013InvestigationOrbits}. Many organisations are also interested in the mitigation of the dangers asteroids pose in the event of a collision with the Earth, a field known as planetary protection \citep{Vardaxis2014Near-EarthDefense, Vardaxis2016ImpactPDC, Lubin2016DirectedDefense, Foster2013MissionTractors}. The development of vehicles for deflecting asteroids on a collision course with the Earth is currently ongoing for such an event, and asteroid exploration provides both technology demonstrations and crucial information as to the material properties of potential threats \cite{Lu2005GravitationalAsteroids}. Further into the future, some organisations are also planning to mine the resources contained within asteroids for sale on terrestrial commodity markets \cite{Elvis2012LetsProfit}.

However, the mass of samples that has been retrieved to date is small, and many scientific investigations require in-situ experiments \citep{Urrutxua2014WhatRH120}. The large distance between Earth and many asteroids further increases mission duration and costs. This is why asteroid retrieval missions \changedTwo{ have been proposed, which aim to place a NEO into an orbit which keeps it close to the Earth \citep{Ceriotti2016ControlOrbitsb}. Future space vehicles and space missions then have easier access to the asteroid's resources \citep{Neves2019MultifidelityAsteroids}.}
Deflection methods -- heavily investigated for use in planetary protection -- could also be used for asteroid retrieval. Methods such as the ion-beam shepherd \citep{Bombardelli2013b}, and the gravity tractor \citep{Lu2005GravitationalAsteroids} have been shown to be effective concepts to facilitate this exploration. 

Given the masses of these asteroids, which are typically on the order of $10^3$ kg, this retrieval is difficult for even moderate retrieval $\Delta v$.
Retrieval must, therefore, be into orbits which can be reached with minimal transfer velocity, termed `low-energy transfers' \citep{Topputo2005LowProblem}.
Often, these transfers exploit the underlying dynamics of the system \citep{Ren2012}. 

\begin{figure}[t]
    \centering
    \includegraphics[width=\linewidth]{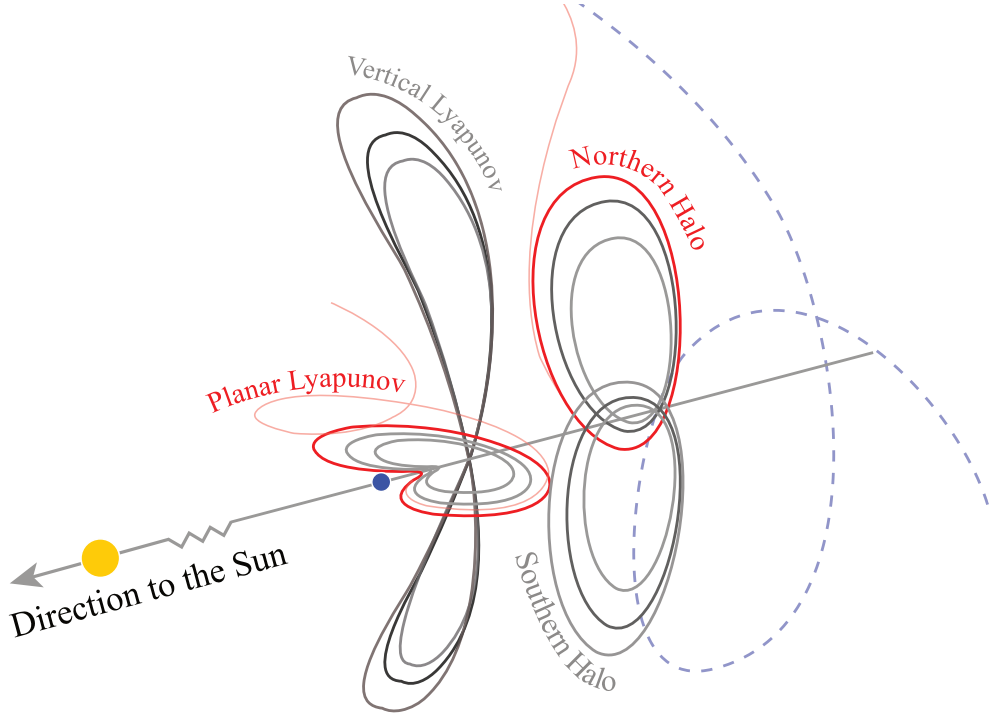}
    \caption{Families of orbits about the $L_2$ point of the Sun-Earth system considered in this work: planar and vertical Lyapunov, and Northern and Southern periodic `Halo' orbits. Highlighted in red are portions of the two retrieval trajectories for asteroids 2020 DW (planar Lyapunov) and 2006 RH120 (Northern Halo), and the nominal trajectory for 2006 RH120 (dashed blue).}
    \label{f:orbit_types}
\end{figure}

A large portion of research has studied the invariant manifolds of the \changedTwo{Sun-Earth} Circular-Restricted Three-body Problem (CR3BP) to facilitate asteroid retrieval, using the periodic orbits about the equilibria of the \changedTwo{system} as final destinations for retrieved the NEOs (Figure \ref{f:orbit_types}) \citep{Mingotti2014CombinedProblem}.
In particular, \citet{Sanchez2012AssessmentResources, GarciaYarnoz2013EasilyPopulation, Sanchez2016} used the Sun-Earth CR3BP to identify a new class of `Easily Retrievable Objects' (EROs), a subset of NEOs which may be retrieved into periodic orbits about the Sun-Earth $L_1$ or $L_2$ points using the invariant manifolds of the CR3BP for a total retrieval cost of less than $500$ m/s. They find $17$ EROs among the approximately $16000$ NEOs known at the time.

\changedTwo{We extend the previous literature in the field by first adjusting parameters used in the identification of the family of EROs, following analysis of the methods and results found in previous work. We go on to construct and study the Pareto fronts for these EROs for the first time through an extensive global optimisation procedure, trading off transfer time and transfer cost.} This analysis shows that the transfer cost is approximately constant for all values of transfer time for the majority of EROs. This extends to zero-time transfers, where many EROs can be captured with a single impulsive manoeuvre at only marginally higher cost than the optimal two-impulse transfer. These discoveries greatly increase flexibility for mission designers. We also find $27$ more EROs and significantly improve on previous solutions, including $17$ new capture trajectories with a transfer $\Delta v$ of less than $100$ m/s. 

This paper is organised as follows: Section \ref{sec:threebodyproblem} \changedOne{reviews} the CR3BP, and Sections \ref{sec:manifolddatageneration}--\ref{sec:candidatedatageneration} describe how we generate the data required for the subsequent optimisation algorithm. Sections \ref{sec:prefilter_method} and \ref{sec:optimisation_method} set out our pre-filtering and optimisation methods, respectively, before the results are discussed in Sections \ref{sec:prefilter_results}--\ref{sec:optimisation_results}. Lastly, we examine the Pareto Fronts of the EROs in Section \ref{sec:pareto_front_results}.

\section{The Circular-Restricted Three-Body Problem}\label{sec:threebodyproblem}

The Circular-Restricted Three-Body Problem (CR3BP) studies the motion of an object of negligible mass under the influence of two far larger bodies, known as primaries, which orbit each other in circles. The CR3BP admits a system of ordinary differential equations to track the motion of the massless object \citep{Szebehely1968}:

\begin{align}
    \ddot{x} &= 2\dot{y} + \frac{\partial\Omega}{\partial x}\\
    \ddot{y} &= -2\dot{x} + \frac{\partial\Omega}{\partial y}\\
    \ddot{z} &= \frac{\partial\Omega}{\partial z}\label{eq:cr3bp_end}
\end{align}{}
where the pseudo-potential function $\Omega$ is defined as:

\begin{equation}
    \Omega = \frac{x^2+y^2}{2} + \frac{1-\mu}{r_1} + \frac{\mu}{r_2}
\end{equation}{}
and $r_{1} = \sqrt[]{(x+\mu)^2 + y^2 + z^2}$, $r_{2} = \sqrt{(x-1+\mu)^2+y^2+z^2}$. The parameter $\mu$ is the \changedOne{dimensionless} mass parameter, which characterises the system.

\begin{equation}
\mu = \frac{m_2}{m_1+m_2}
\end{equation}
where  $m_1$ and $m_2$ are the masses of the two primaries, $m_1 \geq m_2$. These equations are expressed in a synodic (rotating) \changedOne{coordinate} system centred at the barycentre of the primaries, and with the positions of the primaries fixed on the $x$-axis. The distance between the primaries and the gravitational parameter are normalised to unity, and the dimensionless time $t$ is scaled such that the period of the primaries is 2$\pi$. It follows that the mean motion $n$ is unity. For this paper, the two larger bodies are the Sun and the Earth, neglecting the mass of the Moon, such that $\mu$ is $3.0032080443\times 10^{-6}$ throughout the rest of this paper.

The system admits one constant of motion, known as the Jacobi constant $J$, \changedTwo{which is equal to twice the difference between the potential and kinetic energies.} \changedOne{As a result of the non-dimensionalisation of the CR3BP system given above, $J$ itself is dimensionless}, \changedTwo{and the kinetic energy is equal to the square of the scalar velocity $V$, such that:} 

\begin{equation}\label{eq:jacobiconstant}
J = 2\Omega - V^2.
\end{equation}
The Jacobi constant may be considered as analogous to energy in the system.

Five equilibria of the CR3BP exist, known as the Lagrangian points and numbered $L_1$ through $L_5$. Only the $L_1$ and $L_2$ points are considered in this study due to their proximity to the Earth, and their use in previous literature. 

The dynamics about these points is well-known, and bound, periodic orbits exist about them that may be categorised into various families \citep{LlibreMartinezT2001DynamicsPoints}. 
Each of these periodic orbits admits invariant manifolds, with the stable invariant manifold formed by the infinite set of trajectories which asymptotically approach the orbit as $t\rightarrow\infty$, and the unstable invariant manifold formed by the infinite set of trajectories which asymptotically approach the orbit as $t\rightarrow-\infty$.
The stable manifold will thus transport any object onto a periodic orbit solely through the natural dynamics of the system.

\subsection{Reference Frames}

Two reference frames are utilised in this study: the dimensionless synodic frame of the Sun-Earth CR3BP, discussed above, and a dimensional Sun-centred inertial frame in which the states of the NEOs are defined. To simplify interaction with the JPL HORIZONS NEO database used later in this study, time is measured in ephemeris seconds -- seconds past an epoch without leap seconds -- in the Sun-centred inertial frame, with an epoch of J2000, January 1\textsuperscript{st} 2000 12:00. A coordinate transform is required to transform position and velocity vectors from the Sun-Earth CR3BP to the inertial frame for use in later analysis, and is discussed here.

A state with position vector \changedTwo{$\vec{r}_\text{syn}$ and velocity vector $\vec{v}_\text{syn}$ in the non-dimensional synodic system is first rotated to its equivalent state in an inertial frame centred on the barycentre of the primaries $\vec{r}_\text{BCI},~\vec{v}_\text{BCI}$ with dimensionless position and velocity using:}
\begin{align}
    \vec{r}_\text{BCI} &= 
    \mathbf{T}_{IR}~\vec{r}_\text{syn}\\
    \vec{v}_\text{BCI} &= 
    \dot{\mathbf{T}}_{IR}~\vec{r}_\text{syn} + \mathbf{T}_{IR}~\vec{v}_\text{syn}
\end{align}
where the rotation matrices $\mathbf{T}_{IR}$ and $\dot{\mathbf{T}}_{IR}$ are given by
\begin{align}
    \mathbf{T}_{IR} &=
    \begin{bmatrix}
    \cos\theta & \sin\theta & 0 \\ -\sin\theta & \cos\theta & 0 \\
    0 & 0 & 1 \\
    \end{bmatrix} \\
    \dot{\mathbf{T}}_{IR} &= n
    \begin{bmatrix}
    -\sin\theta & \cos\theta & 0 \\
    -\cos\theta & -\sin\theta & 0 \\
    0 & 0 & 0
    \end{bmatrix}
\end{align}
and the rotation angle $\theta$ is given by $\theta = \theta_0 + nt$, where $\theta_0\approx 100.378^\circ$ is the angle subtended between the Earth and the $+x$ axis of the \changedOne{barycentre-centred} inertial frame, measured anticlockwise at epoch. \changedOne{The variable} $t$ is time past the epoch in the non-dimensional time unit of the synodic frame, and $n$ is the mean motion, set to unity.

A translation is applied to move the origin of the system from the barycentre of the two primaries to the centre of the Sun. A translation of $+\mu$ in the $x-$direction completes the transformation from the non-dimensional synodic frame centred at the barycentre of $m_1$ and $m_2$ to a Sun-centred inertial frame with a non-dimensional position and velocity. Equations \ref{eq:r_dimensionalise}--\ref{eq:v_dimensionalise} are then used to dimensionalise the inertial, Sun-centred non-dimensional position and velocity $\vec{r}_\text{dimensionless}$ and $\vec{v}_\text{dimensionless}$ into units of kilometres and kilometres per second, respectively:
\begin{align}
    \label{eq:r_dimensionalise} \vec{r}_\text{dimensionalised} &= \vec{r}_\text{dimensionless} \cdot \text{AU} \\
    \label{eq:v_dimensionalise} \vec{v}_\text{dimensionalised} &= \vec{v}_\text{dimensionless} \cdot \frac{2\cdot\text{AU} \cdot\pi}{\SI{365.26}{\day} \cdot\SI{86400}{\second\per\day}}
\end{align}
with $\text{AU}=1.4959787070\times10^8$ \changedOne{km} the astronomical unit\changedOne{, and $s$ and $d$ the units of seconds and days, respectively.}
\section{Retrieving Near-Earth Objects}\label{sec:datageneration}

The method we use to determine EROs is a three-step `pipeline' that largely follows that of \citep{Sanchez2016} with only minor alterations. \changedTwo{The methods and its alterations are given here as a high-level overview, before being introduced in more detail in following subsections.}

Firstly, the stable hyperbolic invariant manifolds associated with periodic orbits of interest are computed. The orbits chosen here are the planar and vertical Lyapunov orbits, and Northern and Southern `Halo' orbits, to be consistent with previous studies (Figure \ref{f:orbit_types}). A current snapshot of the NEO database and their ephemeris is downloaded from the JPL HORIZONS system provided by the NASA Jet Propulsion Laboratory's Solar System Dynamics Group \citep{NASAJetPropulsionLaboratorySolarSystemDynamicsGroup2020JPLSystem}.

\begin{figure}
\centering
\includegraphics[height=.4\textheight]{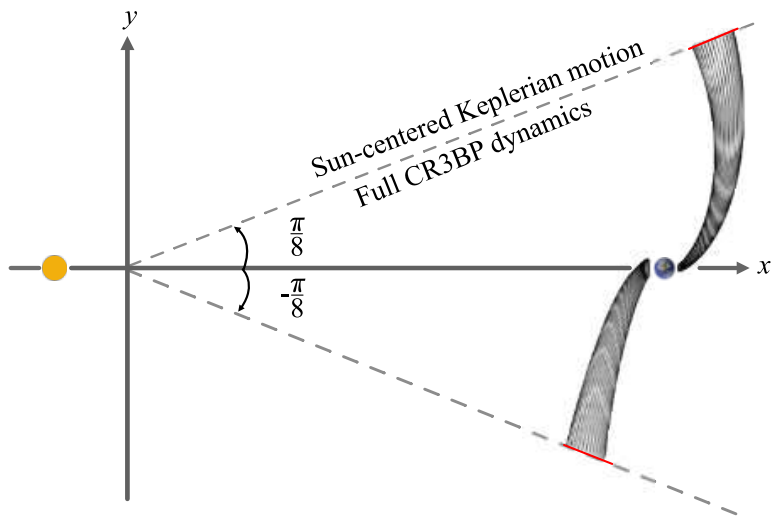}
\caption{The $\pm\pi/8$ planes used as the reference fixed section in this paper: $+\pi/8$ is for manifolds belonging to orbits about $L_2$ and $-\pi/8$ is for orbits about $L_1$. Outside of the `cone' formed by the two planes, the dynamics are assumed to be two-body only around the Sun; inside of the cone, full CR3BP dynamics are used to obtain the stable manifolds.}
\label{f:referenceplane}
\end{figure}

The trajectories that form the manifolds are then integrated backwards in time to their intersection with a plane forming a $\pi/8$ angle with the $x$-axis of the synodic frame of the CR3BP. \changedOne{This integration is discussed in Section \ref{sec:manifolddatageneration}}. Outside of the `cone' formed by the $\pm\pi/8$ planes about the $x$-axis the dynamics are dominated by the Sun (Figure \ref{f:referenceplane}). \changedTwo{This is again in line with previous studies.}

A fast approximator of the transfer cost is then used to pre-filter the NEO catalogue to obtain retrieval candidates. 
Assuming the dynamics to be purely Keplerian and the semi-major axes of their ellipses to be aligned, analytical Hohmann transfers between osculating NEO orbits and intersection points of the manifold with the $\pm\pi/8$ plane are used to estimate the transfer cost.
Any transfer below a given pre-filter threshold qualifies a NEO as a retrieval candidate. \changedTwo{We begin to extend the previous literature at this point: the pre-filter threshold used is investigated and subsequently modified in this work compared to the previous literature, based on this investigation.}

Finally, and again assuming purely Keplerian dynamics, two-impulse Lambert transfers are calculated between the retrieval candidate's osculating orbit at a departure epoch and a point on an invariant manifold outside of the $\pi/8$ cone. The departure epoch, transfer time $t_t$, the target manifold and the point on the manifold are all varied in an optimisation procedure to achieve a minimum capture $\Delta v$. As in the literature, any candidate with a capture $\Delta v$ of less than $500$ m/s is classified as an ERO. \changedTwo{As a modification from the literature, optimisation constraints on the departure epoch are lifted, the use of this $500$ m/s threshold is examined, and the optimisation is extended significantly to study the Pareto fronts of the EROs found.}

\subsection{Orbit and manifold generation}\label{sec:manifolddatageneration}

Families of vertical and planar Lyapunov, and Northern and Southern `Halo' orbits are generated \textit{a priori} within specific ranges of \changedTwo{Jacobi constant}.
The ranges of Jacobi \changedTwo{constant} used are consistent with previous literature to provide a like-for-like comparison. Table \ref{tbl:energy_ranges} provides the ranges used.

\begin{table}
    \centering
    \begin{tabular}{c|c} \toprule
        \textbf{Orbit Family} & \textbf{Jacobi \changedTwo{Constant} Ranges} \\\midrule
        $L_1$ Planar & $\left[3.0003,~3.00087\right]$ \\
        $L_2$ Planar & $\left[2.99985,~3.00087\right]$ \\
        $L_1$ Halo & $\left[3.00042,~3.00082\right]$ \\
        $L_2$ Halo & $\left[3.00025,~3.00082\right]$ \\
        $L_1$ Vertical Lyapunov & $\left[3.0002, 3.00087\right]$ \\
        $L_2$ Vertical Lyapunov & $\left[2.99935,~3.00087\right]$\\
        \bottomrule
    \end{tabular}
    \caption{Ranges of \changedOne{dimensionless} Jacobi \changedTwo{constant} for which orbits were generated, inclusive. $1000$ orbits were generated per family with \changedOne{equispaced} x-coordinates between the extrema given above. The ranges for Halo orbits extend to both the Northern and Southern types.}
    \label{tbl:energy_ranges}
\end{table}

Periodic orbits were found by first correcting an initial guess generated following \citet{Richardson1980} in the full CR3BP dynamics using differential correction. Successive periodic orbits in each family were then found by incrementing the initial $x-$coordinate of each orbit from the previous value. The increment was chosen to yield $1000$ periodic orbits per family between the \changedTwo{Jacobi constant} extrema listed in Table \ref{tbl:energy_ranges}. Individual orbits are identified using a unique index $K$ across all $8$ families, such that $1 \leqslant K \leqslant 8000$.

$360$ points per periodic orbit are used to seed the structure of the stable invariant manifold. For a periodic orbit with an orbital period $\omega$, the seed points are located at regular time intervals of $\omega/360$ around the periodic orbit. These points are assigned an index $n_\text{mnfd}$ for each periodic orbit, with $1 \leqslant n_\text{mnfd} \leqslant 360$.

The points on the periodic orbits are then perturbed along the direction of the stable eigenvalue of the monodromy matrix and integrated backwards in time to the fixed $\pi/8$ section to approximate the intersection of the manifolds with that plane. The $+\pi/8$ plane is used for manifolds belonging to orbits about $L_2$ and the $-\pi/8$ plane is used for manifolds belonging to orbits about $L_1$.  \changedOne{The numerical integration is performed with a 7\textsuperscript{th}/8\textsuperscript{th}-order Dormand-Prince numerical integration scheme provided by the Boost C++ library with relative and absolute integration tolerances of $10^{-13}$.}
In total, $2,880,000$ points on the $\pm\pi/8$ planes are generated for $8000$ orbits.

The intersection points are transformed from the synodic frame of the CR3BP into the Sun-centred inertial frame to simplify further analysis.

\subsection{Candidate data generation}\label{sec:candidatedatageneration}

The ephemeris for each of the $22,406$ NEOs found in the Minor Bodies Database as of 24\textsuperscript{th} February 2020 was downloaded from the NASA HORIZONS system by automating the telnet and FTP interactions with the HORIZONS interface.

The ephemeris were requested between calendar dates Jan 1, 2025 and Jan 1, 2100, with the default time-step. The \changedOne{coordinate} system selected was the ecliptic J2000 frame, centred on the Sun.
The downloads were performed in accordance with the access limits for JPL HORIZONS.
\subsection{Methods for filtering unsuitable NEOs}\label{sec:prefilter_method}

\begin{figure}
    \centering
    \includegraphics[height=.4\textheight]{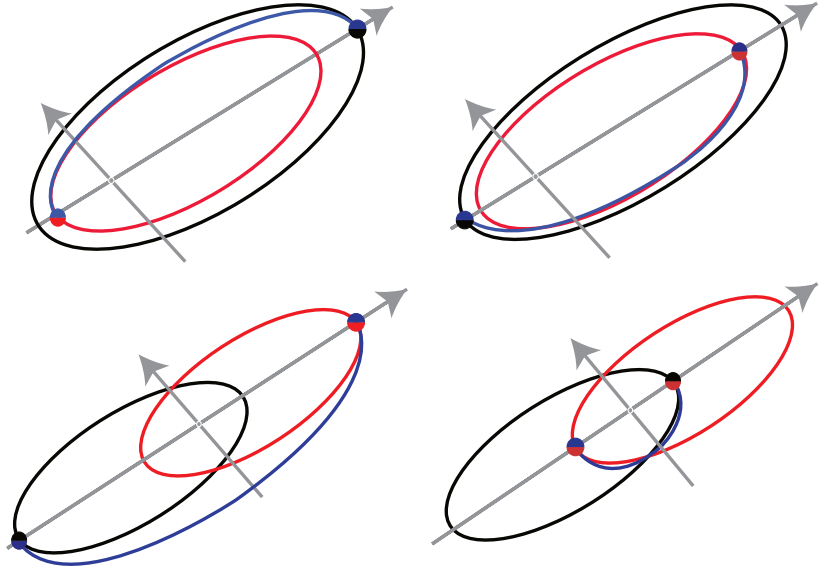}
    \caption{Hohmann transfers considered in the pre-filter. The \changedOne{red orbit is a NEO orbit, the black orbit is the orbit defined by the equivalent Keplerian elements of the manifold target point on the $\pi/8$ plane, and the blue arc is the Hohmann transfer}. Since the orientation of the semimajor axes is ignored, four possible transfers exist. The inclination change is always performed at the furthest point. }
    \label{f:prefilter_method}
\end{figure}

It is unnecessary and computationally infeasible to perform a full optimisation procedure on every NEO included in the Minor Bodies Database. 
Therefore, the Database is filtered using a fast approximator to the optimal transfer cost, based on an idealised Hohmann transfer manoeuvre between a given NEO and the equivalent Keplerian elements of each of the manifold target points. This method greatly simplifies the underlying orbital transfers but is sufficient for rapidly reducing a list of ~$10^4$ NEOs to a list of ~$10^2$ retrieval candidates.

The pre-filter method is outlined in the following and largely follows the same procedure as \citet{Sanchez2016}. However, we use a higher threshold of $3$ km/s to classify a NEO as a retrieval candidate to ensure that no EROs are excluded prematurely, and also use the high-precision ephemerides obtained earlier rather than pre-tabulated orbital elements for the positions of the NEOs.

We assume that the aphelion and perihelion of the NEO and target manifold orbit lie on the line of nodes, and we ignore the relative orientation of the semi-major axes. The cost of matching the aphelion and perihelion of the NEO's orbit and the target manifold point orbit is calculated first:
\begin{align}
\Delta v_{a,~1} &= \sqrt[]{\mu_s \left(\frac{2}{r} - \frac{1}{a_\text{int}}\right)} - \sqrt[]{\mu_s\left( \frac{2}{r} - \frac{1}{a_0}\right)}\\
\Delta v_{a,~2} &= \sqrt[]{\mu_s \left(\frac{2}{r} - \frac{1}{a_\text{f}}\right)} - \sqrt[]{\mu_s\left( \frac{2}{r} - \frac{1}{a_\text{int}}\right)} \\
a_\text{int} &= \frac{a_0 + a_f}{2}
\end{align}
with $\mu_s=1.32712440018\times10^{11}$ \changedOne{km\textsuperscript{3}/s\textsuperscript{2}} the Sun's gravitational parameter, $r$ the radial distance to the Sun at which the transfer is made, and $a_0$, $a_\text{int}$ and $a_f$ the initial, intermediate and final semi-major axes respectively. \changedOne{The radial distance to the Sun and the semi-major axes are in units of km, such that the Hohmann transfer provides approximations to the transfer cost in units of km/s.}
Differences in inclination between the initial and target orbit are then considered by performing an inclination manoeuvre $\Delta v_{i}$ through the angle $|i_f-i_0|$:
\begin{equation}
\Delta v_i = 2~\sqrt[]{\frac{\mu_s}{a}r^*} \sin{\left(|i_f - i_0|/2\right)}
\end{equation}
where $i_0$ and $i_f$ are the initial and final orbit inclinations, respectively. The variable $r^*$ is the ratio of aphelion distance to perihelion distance if the inclination change occurs at perihelion, and its inverse if the change occurs at aphelion. The variable $a$ is the semi-major axis of the orbit whose inclination is being changed, \changedOne{with all units consistent with those given above}.

The total velocity increment $\Delta v_t$ may then be calculated as
\begin{equation}
    \Delta v_t = \sqrt[]{\Delta v_{a,~1}^2 + \Delta v_{i,~1}^2} + \sqrt{\Delta v_{a,~2}^2 + \Delta v_{i,~2}^2}.
\end{equation}

The inclination change is only performed once, such that one of $\Delta v_{i,~1}, \Delta v_{i,~2}$ is zero. 
In total, there are $8$ possible combinations of transfer: $4$ combinations of orienting the semi-major axes of the NEO and the target manifold point, and $2$ combinations per orientation corresponding to whether the inclination change occurs at the beginning or end of the intermediate transfer arc. However, one of the locations of the inclination changes can always be identified as being more expensive based on the radial distance to the Sun. This leaves only four transfers per NEO and target manifold to be considered, as shown in Figure \ref{f:prefilter_method}.

Since the Hohmann transfer approximation depends on the semi-major axis, eccentricity and inclination of both the osculating NEO orbit and the target manifold point, there is the possibility that the pre-filter output may vary with time as the NEOs do not move on purely Keplerian orbits. To account for this, we repeat the pre-filter at regular intervals throughout the time period considered. Once every 28 days was experimentally found to be a good trade-off between reducing the computational expense, while maintaining the fidelity of the problem in trial runs of the pre-filtering routines.
\changedTwo{We note that, while not considered in their original paper, this time dependency is also acknowledged by one of the original authors of \citet{Sanchez2016} in a later publication \citep{Neves2019MultifidelityAsteroids}.}

The pre-filter is implemented in Fortran and interfaced with the SPICE library \citep{Acton1996AncillaryFacility} to access the high-precision ephemerides. To reduce computation time - particularly for when repeating the pre-filter at regular intervals - parallelisation was implemented via the MPI and OpenMP libraries. Approximately one week of CPU time is required to run the asteroid pre-filter at an interval of $28$ days on a 2.0 GHz Intel Xeon E5-2670 processor for the period between January 1, 2025 and January 1, 2100. The time ranges were chosen to match those used in the optimisation procedure (Section \ref{sec:optimisation_method}).

\subsection{Candidate transfer optimisation}\label{sec:optimisation_method}

\begin{table*}[t]
    \centering
    \begin{tabular}{p{.14\textwidth} | p{.49\textwidth} | p{.25\textwidth}}\toprule
    \textbf{Design \newline variable} &  \textbf{Definition} & \textbf{Constraint}\\\midrule
    $t_0$ & The epoch of the first manoeuvre of the transfer. & 2025/01/01 $\leq t_0 \leq $ 2100/01/01\\
    $t_t$ & The duration of the Lambert arc, in days. & $0 \leq t_t \leq 1500$ days.\\
    $t_\text{end}$ & The time in non-dimensional units to arrive at the $\pm \pi / 8$ plane from the manifold insertion point. &  $-25 \leq t_\text{end} \leq 0$\\
    $n_\text{mnfd}$ & Identifies a trajectory within the stable manifold of the periodic orbit, as in Section \ref{sec:manifolddatageneration}. & $1 \leq n_\text{mnfd} \leq 360$\\
    $K$ & Identifies a particular periodic orbit, as in Section \ref{sec:manifolddatageneration}. & $1 \leq K \leq 8000$\\\bottomrule
    \end{tabular}
    \caption{\changedBoth{Definitions of the five design variables used in the optimisation procedure. Taken together, $n_{\text{mnfd}}$ and $t_\text{end}$ uniquely specify the insertion point in the stable manifold of the $K$\textsuperscript{th} target orbit. Dates are given in the form YYYY/MM/DD.}}
    \label{tbl:design_variables}
\end{table*}

%

Two-impulse heliocentric Lambert arcs are optimised between the initial position of the retrieval candidate and a specific point in one of the target manifolds. The epoch of the first impulsive manoeuvre, the transfer duration, and the target point are all varied to achieve insertion onto a manifold with the lowest $\Delta v$. Up to and including $3$ full revolutions are considered for the Lambert arcs.

\begin{figure}
\centering
\includegraphics[width=\linewidth]{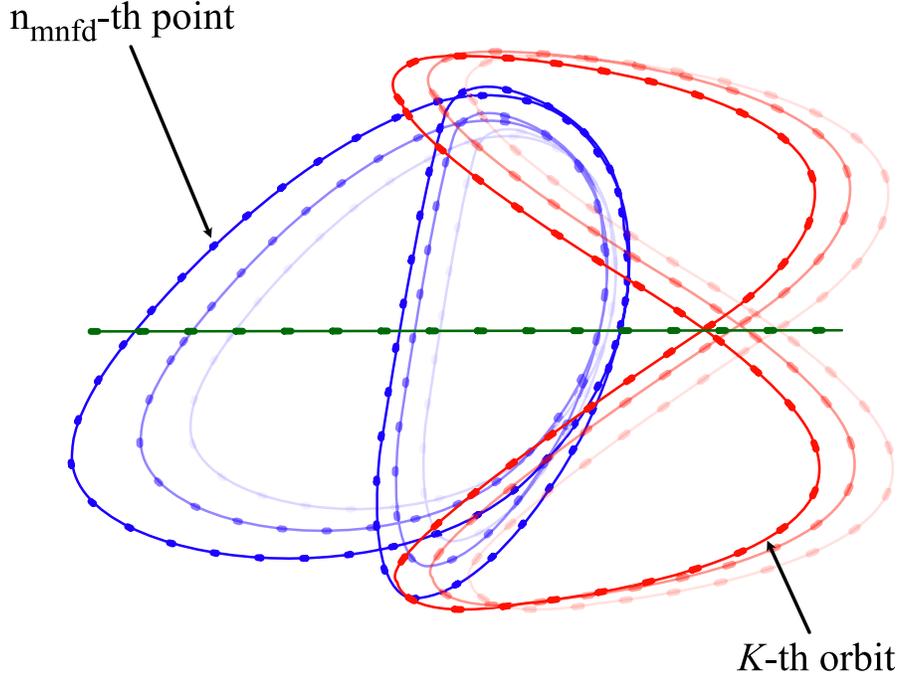}
\caption{
Intersections of stable manifolds and the $\pi/8$ plane for selected periodic orbits from the planar Lyapunov family (green), Northern Halo family (blue), and vertical Lyapunov family (red). Design variables $ 1 \leq n_\text{mnfd} \leq 360$ and  $1 \leq K \leq 8000$ are indicated. }
\label{f:designvariables}
\end{figure}

Five design variables define the optimisation problem: $t_0$, $t_{t}$, $t_\text{end}$, $n_{\text{mnfd}}$, and $K$ (Figure \ref{f:designvariables}). The meaning of these variables, and the constraints placed on them, are given in Table \ref{tbl:design_variables}. 
The constraint on $t_0$ was used in order for it to correspond to the same bounds used in the previous literature, although the lower bound here raised from Jan 1, 2020, for practical reasons. The constraint on $t_t$ was chosen to be any valid transfer length up to the longest transfer found in other literature, at $1500$ days. Similarly, $t_\text{end}$ was constrained to be approximately equal, in non-dimensional units, to the longest `coast' phase after inserting onto the manifold found in other literature.

Non-integer values of $n_\text{mnfd}$ or $K$ chosen by the optimiser are interpolated between the adjacent integer grid points using a cubic B-spline to approximate the manifold between the grid points. Non-zero values of $t_\text{end}$ are numerically integrated backwards from the pre-computed points on the $\pm\pi/8$ plane `\changedOne{on-demand}'.

Two objectives are included in the optimisation problem: the total $\Delta v$ cost of the two-impulse Lambert transfer, and the transfer time $t_t$. This allows us to trade-off transfer time \changedOne{\textit{versus}} cost, as shown in the Pareto fronts in Section \ref{sec:pareto_fronts}.


This optimisation problem is solved using the \changedTwo{off-the-shelf} software MIDACO (Mixed-Integer Distributed Ant Colony Optimiser) \citep{Schlueter2013MIDACOApplications}. MIDACO can perform both global and local optimisation by altering the configuration parameters for the solver. The optimisation was performed using an initial global search, followed by a series of more rigorous optimisations in which the transfer time $t_t$ was kept fixed to generate the Pareto Fronts, as described in Section \ref{sec:pareto_fronts}.

The optimisation is very computationally intensive, and requires dedicated computing facilities to complete in reasonable time. OpenMP and MPI are again used to reduce computation times. The cubic B-spline interpolants are evaluated using the Fortran B-spline library \citep{Williams2019Jacobwilliams/bspline-fortran:6.0.0}, and the Lambert arcs are computed using the Fortran Astrodynamics Toolkit \citep{Williams2020Jacobwilliams/Fortran-Astrodynamics-Toolkit:0.1}. Numerical integration is performed with the 7\textsuperscript{th}/8\textsuperscript{th}-order Dormand-Prince numerical integration scheme provided by the ODEINT C++ library used previously, again interfaced to Fortran. \changedOne{The relative and absolute tolerances in the integration were set to $10^{-13}$}. Approximately $2.5$ days of CPU time is required to run the initial global search on a 2.0 GHz Intel Xeon E5-2670 processor for all the retrieval candidates found by the pre-filter. This corresponds to approximately $10$ minutes of global optimisation per candidate, which was found to be sufficient for MIDACO to identify a suitably close solution to the global minimum.

\subsection{Generating Pareto Fronts for EROs}\label{sec:pareto_fronts}

\begin{figure}
    \centering
    \includegraphics[width=\linewidth]{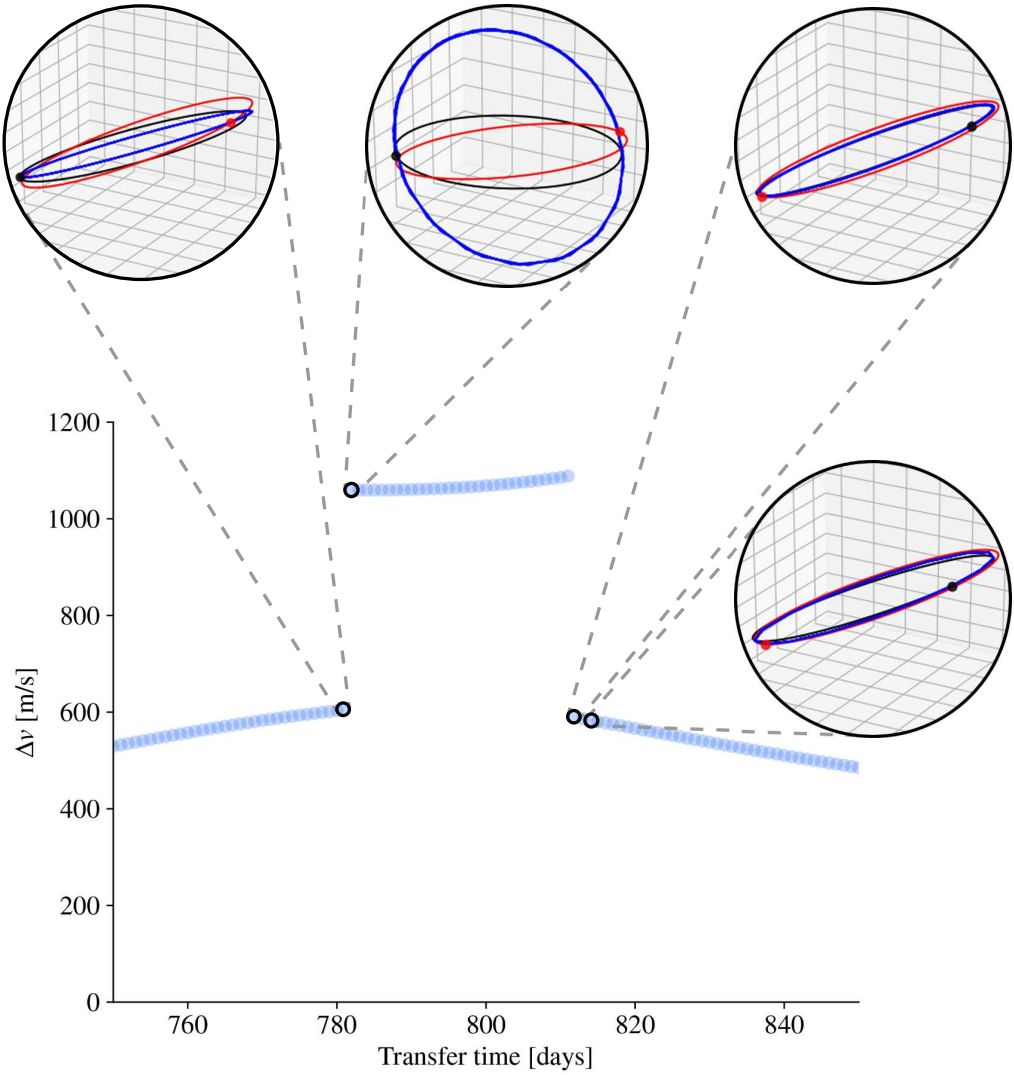}
    \caption{Local $\Delta v$ \textit{vs.} transfer time space for asteroid 2009 BD when the optimal solution at the last transfer time is used as an initial guess. The sudden increases in $\Delta v$ at about $800$ days is a result of the Lambert transfer (blue) between the NEO orbit (red) and equivalent orbit of the target point (black) approaching $\pm 180$ degrees, resulting in transfers requiring a large inclination change.}
    \label{f:preliminary_minima}
\end{figure}

The Pareto Fronts are generated in a further optimisation procedure by discretising the transfer time $t_t$, considering all integer numbers of days between $0$ and $1500$ days. All of the other parameters introduced in Section \ref{sec:optimisation_method} are left \changedOne{unchanged}. Since these solutions may not, therefore, just be continuations of each other but instead widely spread across the search space, careful implementation of the optimisation method is required to find these solutions across the full range of transfer times.

To find these solutions, the Pareto Front optimisation is performed with various initial guesses for each value of $t_t$. One of the initial guesses is the optimal retrieval $\Delta v$ found for the previous value of $t_t$, or the optimal solution found using the method in Section \ref{sec:optimisation_method} when this is unavailable (i.e. for the first iteration). The other guesses are selected randomly to ensure sufficient coverage of the search space.

This strategy derives from analysing the preliminary structure of the local minima for a given NEO: Figure \ref{f:preliminary_minima} shows the structure of the local optima of asteroid 2009 BD between $750$ and $850$ days when only the optimal $\Delta v$ for the previous iterate is used as an initial guess.
There are two distinct sets of behaviour in this Figure. Firstly, there are regions where the retrieval cost is approximately constant, and this is where the optimal solution is merely a continuation of the same family of Lambert arcs with only small changes in the target point. However, in some cases\changedOne{,} there are significant differences in retrieval $\Delta v$ between adjacent values of $t_t$. Viewed over the entire range of $t_t$, these `jumps' are separated by approximately $180$ to $200$ days. This behaviour occurs whenever the Lambert transfer approaches $180$ degrees or every half of an orbital period. For these transfers, the Lambert arc approaches a singularity and small changes in target point yield transfer trajectories with wildly varying inclinations, and thus wildly varying $\Delta v$. This is not inherent to the dynamical system, but an artefact of the definition of the Lambert arc.

Local optimisation is therefore insufficient, and instead a global optimiser must be used to overcome this behaviour and explore the entire parameter space at each value of $t_t$. This is achieved here by providing multiple random initial guesses.
We suspect that a three-impulse transfer with an additional intermediate point would also avoid this artefact.

To save computation time, Pareto Fronts are generated only if the initial global search introduced in Section \ref{sec:optimisation_method} produces an optimal transfer below the ERO threshold. However, the significant time spent in performing the local optimisations to compute these fronts can yield better solutions than the initial global search: solutions are updated when this occurs. 

Successive $10$-minute optimisations for each fixed value of $t_t$ are used \changedOne{to} generate the Pareto Fronts for the EROs identified, requiring approximately \changedOne{$460$} days of CPU time on an Intel Xeon E5-2670 processor. \changedOne{Parallel programming is again used to reduce the wall-time required, and each Intel Xeon E5-2670 CPU has $48$ cores available for this purpose. In total, $144$ cores across three CPUs are used to compute the Pareto fronts in approximately 4 days on the IRIDIS5 High Performance Computing Cluster at the University of Southampton.}

\section{Results}\label{sec:results}

In this section\changedOne{,} we present the results of the algorithm described in the preceding sections.

\subsection{Retrieval Candidates}\label{sec:prefilter_results}

\begin{figure}
    \centering
    \includegraphics[height=.4\textheight]{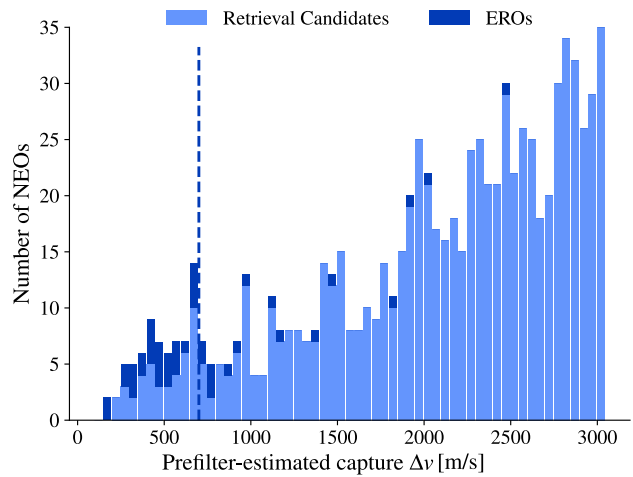}
    \caption{Number of retrieval candidates (light blue) and EROs (dark blue) for a given pre-filter $\Delta v$ threshold. While $70\%$ of EROs are found for $700$ m/s (dashed line), a higher threshold of $3000$ m/s ensures no EROs are excluded prematurely, albeit at a significant computational cost in the pre-filter.}
    \label{f:prefilter_cumulative_count}
\end{figure}

\changedOne{The pre-filter identifies $792$ retrieval candidates when applied to the NEO catalogue.}
The number of candidates that are obtained for a given $\Delta v$ threshold can be seen in Figure \ref{f:prefilter_cumulative_count}: while the number of candidates increases with pre-filter threshold, there are diminishing returns in terms of the number of EROs found. No additional EROs are found between thresholds of $3$ km/s and $5$ km/s, which motivated the use of $3$ km/s as the limit for this work.

Only $46$ of these candidates ($5.8\%$) are retained when lowering the threshold to $0.7$ km/s as in \citet{Sanchez2016}, an increase of $13$ relative to their result.
This increase is to be expected since the NEO database has grown in size by approximately $40$\% since that study. Discarding all candidates discovered after the publication date yields $36$ candidates, an increase of only $3$.

Five candidates that pass the  pre-filter in  the  previous  literature  --  $3332535$, $3339082$, $3520667$, $3634612$, and $3719226$ -- are not recovered here, even when running the pre-filter specifically at the date of the previous study. These missing candidates do not become EROs in the previous literature, so this does not affect the resulting list of EROs.

The time-dependency described in Section \ref{sec:prefilter_method} yields an additional two candidates. This shows that, while a more comprehensive list of candidates can be achieved by considering the changing NEO orbits over time, the significant increase in computing time may not justify the result: a reasonable approximation to the full list of candidates can be found by running the pre-filter just once.
\subsection{Easily Retrievable Objects}\label{sec:optimisation_results}

Of the $792$ retrieval candidates, $44$ go on to become EROs after the full optimisation procedure.
The details of the $\Delta v$-optimal transfers for each ERO identified are given in Table \ref{tbl:ero_results}.
The lowest $\Delta v$ found is $13.97$ m/s and the highest $445.35$ m/s.

\changedOne{Except for} a single ERO, 2011 BL45, we reproduce the full list of EROs in \citet{Sanchez2016}, with an average improvement of $176.7$ m/s ($\sigma = 147.63$ m/s). We also find $17$ new capture trajectories with a $\Delta v$ of less than $100$ m/s. Such objects may be interesting for future study, as they represent significant improvements over the previously-known capture $\Delta v$.

Why 2011 BL45 could not be recovered as an ERO (lowest $\Delta v = 617$ m/s) is not immediately clear, but could be explained by revisions to the orbital data of the NEO since the previous study.

\begin{table*}
\centering
\resizebox{\textwidth}{!}{
\begin{tabular}{lccccr}\toprule
\textbf{SPK ID} & \textbf{$\Delta v [m/s]$} & \textbf{Transfer epoch} & \textbf{Transfer end} & \textbf{Arrival at $\pm\frac{\pi}{8}$ plane} & \textbf{Orbit}\\ 
 \midrule 
3005816 & 162.60 & 2084/12/21 20:46:39 & 2085/07/06 20:46:39 & 2086/02/01 10:31:22 & L1V\\ 
3054374 & 27.39 & 2029/02/04 02:42:50 & 2030/10/05 02:42:50 & 2032/08/04 16:14:46 & L1N\\ 
3076774 & 41.60 & 2091/09/26 03:41:05 & 2093/06/06 03:41:05 & 2095/07/07 12:39:19 & L1V\\ 
3315004 & 27.55 & 2074/09/30 11:54:17 & 2075/03/22 11:54:17 & 2078/05/17 22:15:21 & L1S\\ 
3390109 & 166.09 & 2090/01/13 14:48:27 & 2091/02/27 14:48:27 & 2092/02/19 22:39:48 & L2V\\ 
3403148 & 63.83 & 2025/03/19 08:05:03 & 2027/03/11 08:05:03 & 2027/03/13 20:43:40 & L2N\\ 
3405189 & 50.94 & 2087/10/22 15:41:02 & 2091/06/26 15:41:02 & 2093/01/24 07:46:57 & L1N\\ 
3405338 & 13.97 & 2083/08/31 16:46:35 & 2086/11/12 16:46:35 & 2089/06/27 14:38:03 & L1S\\ 
3435539 & 42.23 & 2029/06/16 05:53:13 & 2030/07/20 05:53:13 & 2030/08/04 15:16:37 & L1V\\ 
3444297 & 136.52 & 2074/07/15 01:23:25 & 2076/10/29 01:23:25 & 2080/04/03 12:48:42 & L1N\\ 
3519518 & 140.95 & 2057/09/13 00:43:54 & 2060/12/01 00:43:54 & 2061/09/02 11:08:22 & L1P\\ 
3549639 & 146.66 & 2092/11/17 04:32:51 & 2095/12/09 04:32:51 & 2098/08/30 16:03:43 & L2V \\
3550232 & 195.40 & 2059/10/03 05:07:28 & 2062/03/25 05:07:28 & 2065/01/24 03:42:44 & L1N\\ 
3551168 & 180.83 & 2035/02/16 23:00:39 & 2036/08/25 23:00:39 & 2037/05/30 06:05:40 & L2V\\ 
3556127 & 169.01 & 2073/01/15 19:47:19 & 2073/10/30 19:47:19 & 2073/11/24 15:04:24 & L2V\\ 
3568303 & 418.53 & 2063/03/19 15:57:36 & 2065/12/08 15:57:36 & 2067/12/16 17:26:46 & L2V \\
3582145 & 111.98 & 2039/11/11 04:56:22 & 2040/07/18 04:56:22 & 2040/07/18 08:34:39 & L1V\\ 
3610175 & 90.82 & 2097/06/19 02:55:06 & 2100/12/02 02:55:06 & 2103/11/20 14:56:24 & L2V\\ 
3618493 & 140.38 & 2085/10/31 13:57:51 & 2089/08/16 13:57:51 & 2090/11/08 03:07:06 & L1S\\ 
3625129 & 203.01 & 2090/12/27 06:25:33 & 2092/02/26 06:25:33 & 2092/05/11 15:40:21 & L1S\\ 
3648046 & 355.29 & 2041/11/26 12:08:11 & 2045/09/23 12:08:11 & 2049/04/05 01:23:13 & L2V\\
3696401 & 69.84 & 2073/11/14 22:32:44 & 2076/03/14 22:32:44 & 2079/12/15 16:03:31 & L1N\\ 
3697803 & 115.13 & 2038/07/11 09:56:29 & 2040/03/31 09:56:29 & 2041/03/09 15:04:05 & L1V\\ 
3698849 & 23.89 & 2091/05/27 20:52:28 & 2095/07/05 20:52:28 & 2096/02/06 05:44:02 & L1S\\ 
3702319 & 56.93 & 2080/12/13 23:13:48 & 2084/01/05 23:13:48 & 2085/03/28 03:06:42 & L1V\\ 
3719859 & 445.35 & 2036/03/14 17:07:21 & 2039/03/02 17:07:21 & 2040/11/08 13:58:20 & L2N\\ 
3726012 & 108.25 & 2047/04/27 02:52:56 & 2051/01/14 02:52:56 & 2053/06/13 02:13:48 & L1N\\ 
3733264 & 23.68 & 2037/10/07 12:46:09 & 2040/08/23 12:46:09 & 2044/02/08 23:50:25 & L1V\\ 
3735181 & 123.71 & 2064/12/07 21:54:01 & 2068/09/14 21:54:01 & 2071/12/06 08:15:26 & L1P\\ 
3745994 & 301.71 & 2030/01/29 13:28:25 & 2033/10/17 13:28:25 & 2035/07/21 02:59:40 & L2S\\ 
3748467 & 332.02 & 2045/07/11 09:31:04 & 2047/07/20 09:31:04 & 2049/07/17 15:15:49 & L1P \\
3759358 & 89.76 & 2048/06/09 21:15:26 & 2051/07/02 21:15:26 & 2052/04/18 09:36:45 & L2N \\
3781977 & 305.79 & 2072/12/16 21:24:12 & 2073/07/23 21:24:12 & 2075/01/11 19:26:12 & L2V\\
3781983 & 245.83 & 2099/06/01 04:29:49 & 2102/07/12 15:22:23 & 2105/07/20 03:40:37 & L1P \\ 
3825158 & 170.93 & 2095/05/25 09:19:32 & 2099/04/20 09:19:32 & 2102/08/26 09:36:23 & L1P \\
3826850 & 254.37 & 2069/08/12 23:03:50 & 2071/04/19 23:03:50 & 2074/05/24 00:54:36 & L2V\\ 
3836857 & 51.52 & 2049/01/24 00:51:25 & 2051/10/17 00:51:25 & 2052/11/10 00:13:14 & L2V\\ 
3840784 & 81.52 & 2083/02/05 15:22:51 & 2086/04/24 15:22:51 & 2089/10/10 05:13:44 & L1V\\ 
3843512 & 33.06 & 2087/07/16 05:05:26 & 2090/04/14 05:05:26 & 2092/06/05 09:31:46 & L1V\\ 
3843865 & 15.25 & 2093/06/22 11:37:34 & 2096/06/08 11:37:34 & 2096/08/01 04:44:51 & L2S\\ 
3879383 & 128.16 & 2051/07/27 05:54:09 & 2053/12/07 05:54:09 & 2057/07/19 05:53:24 & L1N\\ 
3893945 & 94.96 & 2099/10/21 08:33:59 & 2103/02/09 08:33:59 & 2106/12/15 20:49:44 & L1N\\ 
3989308 & 116.72 & 2082/08/03 00:48:08 & 2086/07/09 00:48:08 & 2088/11/26 02:02:44 & L2V\\ 
3991650 & 112.27 & 2073/12/06 05:09:00 & 2076/11/03 05:09:00 & 2079/12/29 17:58:33 & L2P \\
\bottomrule\end{tabular}
}
\caption{List of $\Delta v$-optimal solutions found for all of the EROs in this work; we find $17$ new trajectories with retrieval costs under 100 m/s. Note that while these transfers may be optimal in terms of $\Delta v$, many more solutions exist for other transfer times with similar retrieval velocities. All dates are given in the form YYYY/MM/DD HH:MM:SS.}
\label{tbl:ero_results}
\end{table*}

Only $31$ of these EROs are recoverable with a pre-filter threshold of $0.7$ km/s, which suggests that the threshold used in previous literature limits the final number of EROs by 30\%. Note that this still represents an improvement over previous literature of $14$ EROs.

An additional constraint was placed on $t_0$ in the previous literature to reduce the search space and computational cost by limiting $t_0$ to the first astronomical synodic period of the NEO and the Earth. This constraint was found to increase the optimal $\Delta v$ by approximately $50$ m/s, as the behaviour of the NEO is not perfectly periodic between astronomical synodic periods.

\begin{figure}
    \centering
    \includegraphics[height=.4\textheight]{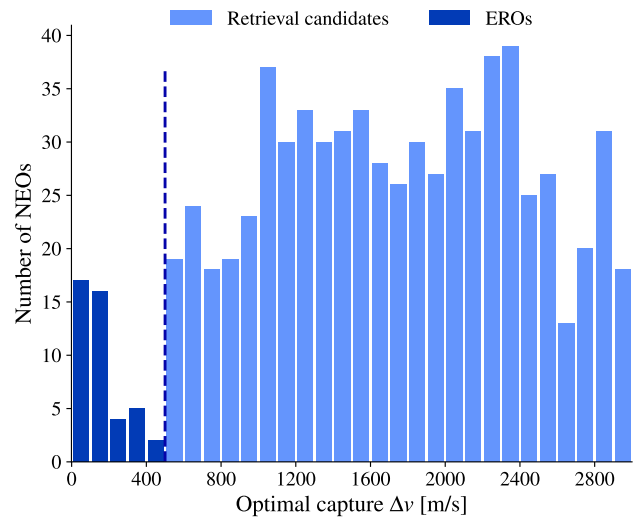}
    \caption{Distribution of retrievable NEOs over the range of optimal capture $\Delta v$. The dark blue bars correspond to EROs, but many more NEOs have very similar capture $\Delta v$. EROs are subject to more rigorous optimisation, causing the gap between dark blue and light blue bars.}
    \label{f:optimiser_dv_pair}
\end{figure}

Attention must be paid to the $500$ m/s threshold used for defining a candidate, which was originally selected in \citet{GarciaYarnoz2013EasilyPopulation}. We are not aware of any obvious engineering constraint to set the ERO threshold at that value. Figure \ref{f:optimiser_dv_pair} shows the number of NEOs in discrete bins of optimal capture $\Delta v$.
By increasing the threshold by as little as 20\%, the number of EROs increases by 50\%. Note that in Figure \ref{f:optimiser_dv_pair}, the dark blue region corresponds to EROs subjected to the more extensive optimisation procedure to generate the Pareto fronts. This explains why these solutions exhibit much lower $\Delta v$ and the apparent gap in the distribution. It is expected that the solutions in the light-blue regions would similarly improve when subjected to the same, more extensive optimisation procedure.

Instead of using an arbitrary fixed cutoff, a mission-by-mission approach could be used. The maximum retrieval $\Delta v$ for a given mission profile may be used in combination with Figures \ref{f:prefilter_cumulative_count}--\ref{f:optimiser_dv_pair} to identify the population of EROs for that mission and available computational resources.
Such an approach would allow a more flexible and practical definition of an ERO: transfer costs of $499$ m/s and $501$ m/s are functionally the same, but the latter is deemed un-retrievable by using a fixed cutoff.

Lastly, all identified transfers are validated in the full CR3BP model through another local optimisation. \changedOne{In general, the transfers in the full CR3BP differ only on the order of m/s. While this may be significant for objects such as $3405448$, at a retrieval cost of $13.97$ m/s, it is only one or two per cent of the ERO classification threshold, and all EROs identified remain so in the full CR3BP dynamics. This further validates the choice of the $\pm\pi/8$ planes in the analysis.}

\subsection{Pareto Fronts for EROs}\label{sec:pareto_front_results}

We find that the optimal $\Delta v$ for typical EROs is not unique. Instead, for a given ERO, many different combinations of transfer time, departure epoch, and target orbit yield very similar capture $\Delta v$ to the optimal solution. In fact, for 28 of the 44 EROs found, more than $95\%$ of their solutions fall below the ERO threshold of $500$ m/s.

This can be explained by the geometric properties of the invariant manifolds used for retrieval.
In backwards time, the structure of the manifold `stretches', and occupies large regions of the Sun-Earth system. Thus, for small changes in departure epoch or transfer time, there exists a range of nearby target points for which a transfer may occur for a comparable $\Delta v$.

\begin{figure*}
\centering
\begin{subfigure}{.75\linewidth}
\centering
\captionsetup{width=\textwidth}
\includegraphics[width=\textwidth]{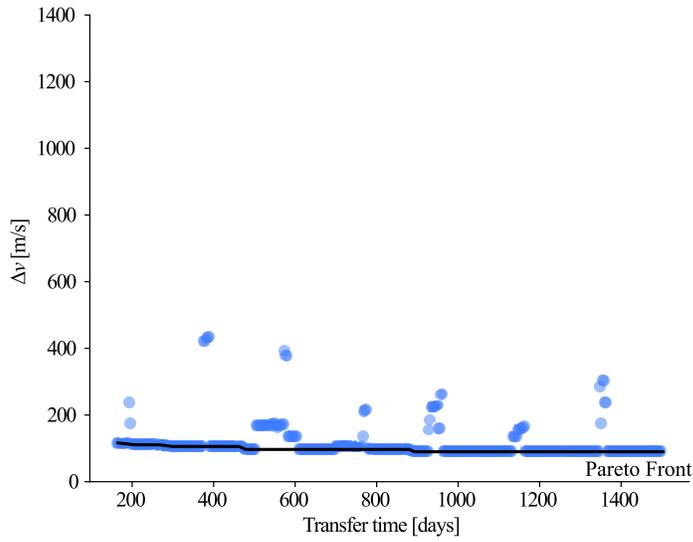}
\caption{Pareto Front for asteroid 2012 TF79, which is an ERO for any transfer time, including a single-impulse, zero-time transfer. The Pareto front is shown as a black line.}
\label{f:pareto_start}
\end{subfigure}
\begin{subfigure}{.75\linewidth}
\centering
\captionsetup{width=\textwidth}
\includegraphics[width=\textwidth]{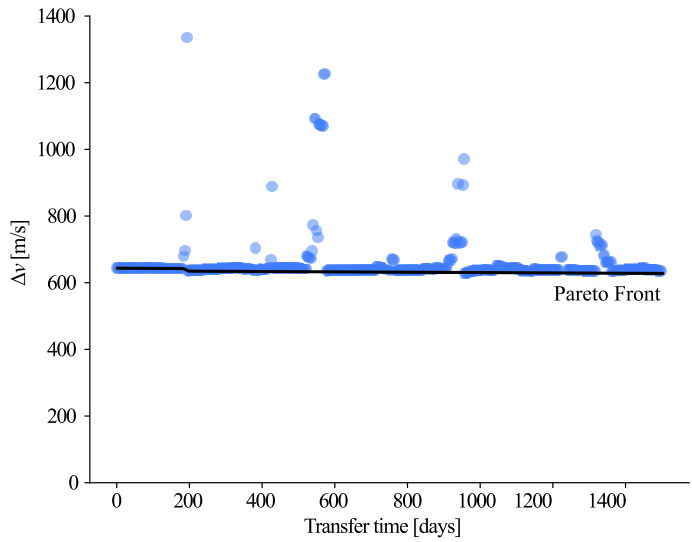}
\caption{Pareto Front for asteroid 2011 BL45. Note that 2011 BL45 is not an ERO, but displays similar behaviour for its Pareto front as EROs, showing that non-EROs may also offer the same flexibility to mission design as EROs.}
\label{f:pareto_end}
\end{subfigure}
\caption{Pareto fronts for asteroids 2012 TF79 (left) and 2011 BL45 (right).}
\end{figure*}

To illustrate this, consider the Pareto front for asteroid 2012 TF79 (Figure \ref{f:pareto_start}). This asteroid admits an optimal-$\Delta v$ retrieval cost of $90.82$ m/s. However, it also exhibits retrieval trajectories below the $500$ m/s threshold for any transfer time within the allowed range.

This behaviour is not just limited to candidates meeting the formal requirement for EROs: 2011 BL45 has an optimal retrieval $\Delta v$ of approximately $617$ m/s, above the threshold for an ERO. But, in this case, for more than $95\%$ of $t_t$ values between $0$ and $1500$ days\changedOne{,} the resulting transfer costs stays below $650$ m/s, quite close to the optimal solution (Figure \ref{f:pareto_end}). 
This shows that there is significant flexibility when designing retrieval missions at the cost of very small changes to the total $\Delta v$.

Candidates which admit more solutions below the ERO threshold may be more promising for future retrieval investigations for space missions since they are more resilient to changes in -- or disruptions to -- mission timelines.
These candidates may also allow for material samples to be returned over larger \changedOne{periods}, since many different low-$\Delta v$ transfers exist. Such a strategy would see samples returned at regular intervals during a long-duration stay and could provide a greater yield from the asteroid without having to capture the entire asteroid.

The last key result we find is the existence of single-impulse transfers for \changedOne{a} similar cost to their optimal two-impulse counterparts for many EROs. In these transfers the transfer time $t_t$ is zero, reducing the Lambert transfer to a single manoeuvre.  
In the cases of 2011 BL45 and 2012 TF79, these objects are retrievable for a single-impulse transfer for only $2.5\%$ and $9\%$ more than their optimal solution, respectively. 

Such single impulse transfers can significantly reduce the complexity of retrieval missions and are therefore of particular interest.
This also shows that the (nearly) single-impulse transfer for asteroid 2006 RH120 highlighted in \citet{Sanchez2016} is \changedOne{actually} a generic trait shared across most EROs.

\section{Conclusion and Future Work}

This paper has outlined methods to determine the population of Easily Retrievable Objects, Near-Earth Objects for which capture trajectories using the invariant manifolds of the Sun-Earth CR3BP exist below a threshold of $500$ m/s.
Limitations in the pre-filter threshold used in previous works have been investigated, and the effect of this on the total number of EROs found has been quantified. We discover $27$ more EROs, including $17$ new capture trajectories below $100$ m/s, and improve on previous solutions by an average of $176.7$ m/s. The Pareto fronts in the $\Delta v$ \changedOne{\textit{versus}} transfer time space have been studied for the first time, and we identify that the optimal transfer cost for these EROs is not unique. A key result is the existence of single-impulse transfers for similar cost to their two-impulse counterparts, which could provide significant flexibility to mission designers.

It is important to note that, while there are generally improvements on the transfers given in previous literature, the transfers often have different transfer epochs and durations. Such a phenomenon may be explained by revisions to the states of the NEOs.

While in this work the ranges of \changedTwo{Jacobi constant} for which target orbits are generated is limited to those in Table \ref{tbl:energy_ranges}, \changedOne{following} the literature \citep{Sanchez2016}, the lower bound is arbitrary and it is possible to extend the lower bound further, corresponding to periodic orbits moving farther away from the Lagrange point. Preliminary tests on the L1 Planar family of periodic orbits indicate that the number of NEOs passing the pre-filter doubles if the lower bound is reduced to $2.99978325$. The effect of this on the final number of EROs will be investigated in future work.

The use of the impulsive transfer approximation is noted as a potential limitation, as the high mass of many of the EROs found makes impulsive transfers physically impossible. However, given that many of the transfers presented here have transfer times on the order of hundreds of days, constant low-thrust propulsion may be able to replace the two impulses for similar $\Delta v$, as done in \citet{Sanchez2016}. 

The authors also believe that more EROs exist among the population of retrieval candidates identified. Given the improvements in solutions observed during the optimisation steps for the Pareto fronts, we expect that some candidates with velocities close to the ERO threshold could yield solutions below the threshold if subjected to further optimisation.

\changedOne{To} facilitate future analysis, all data supporting this study are openly available from the University of Southampton repository at \linebreak\url{https://doi.org/10.5258/SOTON/D1631}.

\section*{Acknowledgements}

The authors would like to acknowledge financial support from the EPSRC Centre for Doctoral Training in Next Generation Computational Modelling grant EP/L015382/1, and the use of the IRIDIS High Performance Computing Facility and associated support services at the University of Southampton.

\appendix

\bibliographystyle{elsarticle-num-names} 
\bibliography{Mendeley}

\end{document}